\documentclass[10pt,a4paper,english,twocolumn]{IEEEtran} 
\usepackage[T1]{fontenc}
\usepackage{color, colortbl}
\usepackage{babel}
\usepackage{verbatim}
\usepackage{textcomp}
\usepackage{amsmath}
\usepackage{amstext}
\usepackage{graphicx}
\usepackage{tabulary}
\usepackage{caption}
\usepackage{float}
\definecolor{Gray}{gray}{0.95}
\definecolor{LightCyan}{rgb}{0.8,0.85,1}
\definecolor{LightBlue}{rgb}{0.6,0.6,1}

\makeatletter

\pdfpageheight\paperheight
\pdfpagewidth\paperwidth

\usepackage{epstopdf}
\usepackage{cite}
\usepackage{balance}

\makeatother

\addto\captionsenglish{}

\begin{document}

\title{IEEE 802.11be Extremely High Throughput: \\ The Next Generation of Wi-Fi Technology \\ Beyond 802.11ax
}

\author{
{David L$\acute{\textrm{o}}$pez-P$\acute{\textrm{e}}$rez}, 
{Adrian Garcia-Rodriguez}, 
{Lorenzo Galati-Giordano}, 
{Mika Kasslin},  and
{Klaus Doppler}
\thanks{The authors are with Nokia Bell Labs.}
\thanks{$\copyright$2019 IEEE. Personal use of this material is permitted. Permission from IEEE  must  be  obtained  for  all  other  uses,  including  reprinting/republishing this material for advertising or promotional purposes, collecting new collected works for resale or redistribution to servers or lists, or reuse of any copyrighted component of this work in other works.}
\thanks{Digital Object Identifier 10.1109/MCOM.001.1900338.}
\thanks{The published version of the article can be found at: https://ieeexplore.ieee.org/document/8847238}
}
\maketitle
\begin{abstract}

Wi-Fi technology is continuously innovating to cater to the growing customer demands,
driven by the digitalisation of everything, 
both in the home as well as the enterprise and hotspot spaces. In this article, we introduce to the wireless community the next generation Wi-Fi---based on IEEE 802.11be Extremely High Throughput (EHT)---,
present the main objectives and timelines of this new 802.11be amendment,
thoroughly describe its main candidate features and enhancements,
and cover the important issue of coexistence with other wireless technologies.
We also provide simulation results to assess the potential throughput gains brought by 802.11be with respect to 802.11ax.

\end{abstract}


\vspace{-0.2cm}
\section{Introduction\label{sec:Introduction}}

Wi-Fi technology is among the greatest success stories of this new technology era, 
and its societal benefits are known to most of the world population. 
Wi-Fi has connected and entertained people,
and has assisted in the creation of new technologies, industries and careers around the globe. 
According to a recent report from the Wi-Fi Alliance~\cite{2018WifiEconomicValue},
more than 9\,billion Wi-Fi devices are currently in use world-wide, 
where individuals, families, governments and global organisations depend on Wi-Fi every day.
According to the same report,
the economic value provided by Wi-Fi reached the astounding amount of nearly \$2\,trillion by 2018, 
and is forecasted to grow to almost \$3.5\,trillion by 2023. 
Since Wi-Fi has become an essential part of the home,
and a key complementary technology for both enterprise and carrier networks, 
this economic value is only expected to increase beyond 2023, 
as the newly defined generation of more capable Wi-Fi products---Wi-Fi\,6, based on the most recent Institute of Electric and Electronic Engineers (IEEE) 802.11ax specification~\cite{2018IeeeStandardAxDraft}---becomes widely available.  

Concurrently, the requirements of wireless data services continue to increase in many scenarios
such as homes, enterprises and hotspots,
beyond the capabilities of Wi-Fi\,6. 
Video traffic will be the dominant traffic type in the years to come,
and its throughput demand will keep growing to tens of Gbps with the emergence of more sophisticated technologies, 
e.g. 4k~\&~8k video. 
Simultaneously, new applications demanding both high-throughput and low-latency are also proliferating. 
Among these, augmented and virtual reality, gaming, remote office and cloud computing pose the stringiest requirements, 
including sub-5\,ms latency needs.
Reliability is also a major concern in the emerging digital industry,  
where guaranteeing that 99.99\% of the data packets are correctly delivered within their given deadline may be the bare minimum to replace wired with wireless communications. 
With these high-throughput, low-latency and high-reliability requirements\footnote{Other quality of service key performance indicators may apply depending of the service nature.}, 
consumers \mbox{will demand a further improved Wi-Fi.}

To address such expectations,
the IEEE 802.11 working group (WG)---responsible for the standardisation of the medium access control (MAC) and physical (PHY) layers of Wi-Fi products---has already initiated discussions on new technical features for bands between 1 and 7.125\,GHz to make sure that Wi-Fi products deliver up to the highest standards.
To this end, 
and more concretely,
the IEEE 802.11 WG approved the formation of a topic interest group (TIG), 
and subsequently, a study group (SG) and a task group (TG) on May 2018, July 2018 and May 2019, respectively. 
These groups had and still have the primary objective of ensuring that the next generation of Wi-Fi---referred to as IEEE 802.11be Extremely High Throughput (EHT)---meets the peak throughput requirements set by upcoming applications~\cite{2018CariouEhtPar}.  


In this article, we introduce the main objectives of 802.11be,
as well as the views from different Wi-Fi stakeholders about the process and timelines to generate such amendment (Sec.~\ref{sec:objective}). 
We dive into the main candidate technical features of 802.11be,
presenting each one of them,
and describing their benefits and challenges (Sec.~\ref{sec:features}).
Moreover, we discuss the important issue of coexistence (Sec.~\ref{sec:coexistence}),
and we provide system-level simulation results that show some of the potential throughput gains that 802.11be may provide (Sec.~\ref{sec:simResults}). 
Altogether, this article aims to become an accessible guide to 802.11be for researchers and the general audience interested in Wi-Fi. 

\begin{figure*}[!t]
\centering
\includegraphics[width=17.5cm]{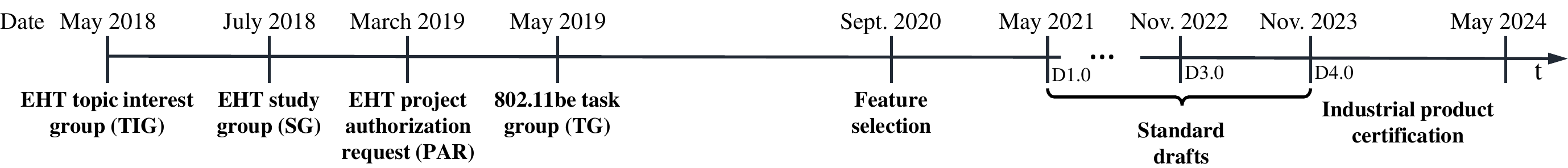}
\caption{Illustration of the initial standardization timeline agreed by 802.11 stakeholders.}
\label{fig:EHTDevelopmentCycle}
\vspace{-0.5cm}
\end{figure*}

\vspace{-0.2cm}
\section{Objectives and timeline \label{sec:objective}}
 
The Wi-Fi community is aiming high,
with the recently established 802.11be (TG)  targeted at
\begin{itemize}
\item[{i)}] 
enabling new MAC and PHY modes of operation capable of supporting a maximum throughput of at least 30\,Gbps, 
measured at the MAC data service access point (AP)---4$\times$ w.r.t. 802.11ax---
using carrier frequencies between 1 and 7.125 GHz, while 
\item[{ii)}] 
ensuring backward compatibility and coexistence with legacy 802.11 devices in the 2.4, 5 and 6\,GHz unlicensed bands~\cite{2018CariouEhtPar}.
\end{itemize}
Moreover, 802.11be will define at least one mode 
of operation capable of improved worst case latency and jitter.\footnote{
It is important to note the 802.11be TG has not defined any specific objectives in terms of latency and/or reliability so far,
and a more detailed analysis on this matter was carried out by the real time application (RTA) TIG \cite{RTAReport}.
}

Traditionally, major 802.11 amendments, 
such as 802.11n or 802.11ax, 
have taken 6+ years to complete, 
with 
\begin{itemize}
\item[{i)}] 
a development process effectively serialised---TG formation, feature set definition, draft development and certification process---, and 
\item[{ii)}] 
no major overlap between subsequent amendments.
\end{itemize}

In the first meeting of the 802.11be TG, held on May 2019, 
the TG members decided to stick to the current---and so far successful---amendment development model but with a shorter time span. 
The approved timeline is at least 6 months faster than that of 802.11ax, 
and is illustrated at the top of Fig.~\ref{fig:EHTDevelopmentCycle}. 
This next major 802.11 amendment is expected to span 5 years and deliver massive---and not just moderate---enhancements on many fronts for 802.11 multitudinous customers and use cases. 
This timeline also enables the Wi-Fi Alliance to develop appropriate certifications of high profile features
in a timely manner to satisfy market needs~\cite{2018HartNoCascading}.

\section{Candidate technical features \label{sec:features}}

A variety of candidate technical features have been proposed by numerous industrial and academic experts in the 802.11be fora. In the following, we describe those that have attracted the most attention.

\vspace{-0.2cm}
\subsection{320\,MHz bandwidth and more efficient utilisation of non-contiguous spectrum \label{sec:320MHz}}

\begin{figure*}[!t]
\centering
\includegraphics[width=18cm]{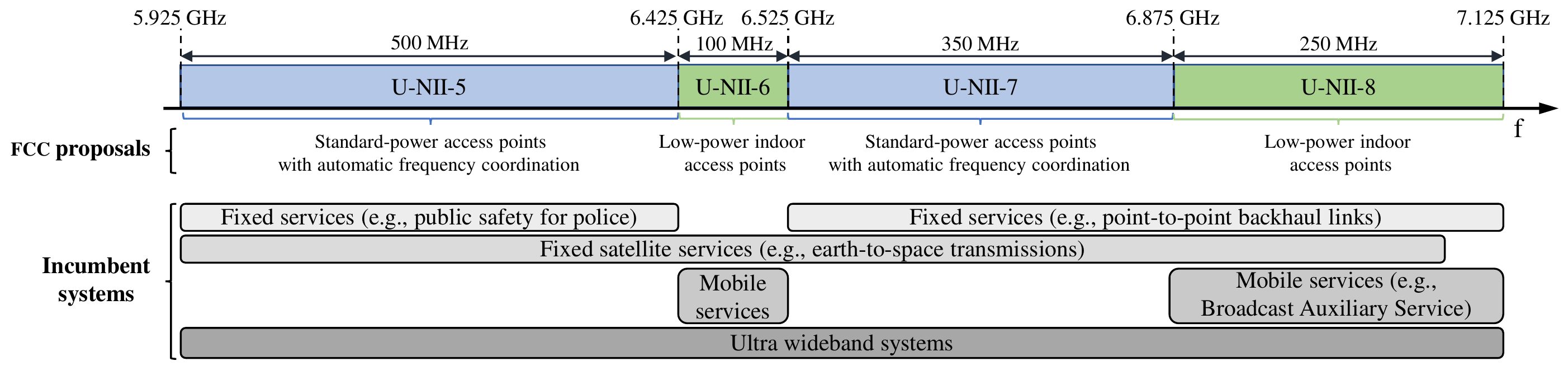}
\caption{Proposed Federal Communications Commission (FCC) frequency allocation in the 6\,GHz band. Please refer to  \cite{FCCUnlicensed6GHz} for further details on nomenclatures, band allocations and proposed coexistence techniques.}
\label{fig:FCCAllocation}
\vspace{-0.5cm}
\end{figure*}

Spectrum is the air that wireless networks breath,
and any new generation of radio technology always attempts to leverage the usage of new spectrum bands,
as they become available. 
802.11be  is no exception, 
and following the initial steps of 802.11ax, 
Wi-Fi stakeholders embrace the usage of the 6\,GHz band 
as an immediate approach to increase Wi-Fi peak throughput, 
as shown in Fig.\ \ref{fig:FCCAllocation}. 
In this regard, discussions about the most efficient approaches to operate the up to 1.2\,GHz of potentially accessible unlicensed spectrum between 5.925 and 7.125\,GHz---which more than doubles the available bandwidth in the 5\,GHz band---are ongoing.

The adoption of 160\,MHz and 320\,MHz communication bandwidth per AP in the 6\,GHz band as mandatory and optional features, respectively, seems a sensible choice,
building on 802.11ax, 
where 160\,MHz bandwidth per AP is already an option~\cite{2018ChenDiscussionPHYFeatures}.
Moreover, a minimum channel size of 40 or even 80\,MHz in the 6\,GHz band also seems appropriate when compared to the 20\,MHz one used in the 2.4 and 5\,GHz bands,
given the focus on extremely high throughput.

While the benefits of using the 6\,GHz band to enhance peak and system throughputs are obvious,
the usage of a new band also opens up the opportunity for new networking approaches.
For example, there are on-going discussions on whether 802.11be-compliant APs should
\emph{i)} always schedule uplink transmissions in the 6\,GHz band---thereby reducing the time spent on channel contention---, and
\emph{ii)} have the capability to request 802.11ax devices to vacate the 6\,GHz band on demand,
to reinforce such coordinated access.

 
\vspace{-0.2cm}
\subsection{Multi-band/multi-channel aggregation and operation \label{sec:multi-channel}}

With the emergence of dual-radio STAs and tri-band APs capable of simultaneously operating at 2.4, 5 and 6\,GHz, 
one of the main objectives of 802.11be is to make a more efficient use of these multiple bands and channels therein. 
We describe four of the most appealing techniques being considered by 802.11be in the following~\cite{2019PoKaiMultiLinkOp}.
\begin{itemize}
\item [\emph{a)}] \emph{Multi-band data aggregation.} 
The aggregation of 5 and 6\,GHz spectrum for data transmission or reception is a feature fully aligned with 802.11be fundamental objective of enhancing Wi-Fi's peak throughput \cite{2019PoKaiMultiLinkOp}. 
Effectively, this aggregation may require Wi-Fi devices to synchronise the start of the transmission opportunity (TXOP) in different bands, 
therefore making this approach more efficient in sparsely populated scenarios 
where contention for channel access is generally smoother.

\item [\emph{b)}] \emph{Simultaneous transmission and reception in different bands/channels.} 
This feature, 
also commonly referred to as multi-band/multi-channel full duplex, 
has the potential of reducing the communication latency and enhancing the throughput by enabling an asynchronous and simultaneous uplink/downlink operation in separate bands/channels \cite{2019PoKaiMultiLinkOp}. If this feature is to be included in 802.11be, 
a minimum separation between the downlink and uplink channels within the same band is likely to be included to prevent uplink to downlink and downlink to uplink interference.

\item [\emph{c)}] \emph{Simultaneous transmission and reception in the same channel.} 
In parallel to the 802.11be TG, 
the 802.11 WG also approved the formation of a TIG in January 2018 to examine the technical feasibility of full duplex operation for Wi-Fi~\cite{2018FDreport}. 
The TIG finished its activity in December 2018, 
concluding that full duplex operations can be realised with minor modifications to the 802.11 standard, 
and can yield various benefits such as increased throughput per STA, reduced latency, collision detection and hidden node mitigation~\cite{2013NextGenWirelessLAN} within a densely populated basic service set. 
Accordingly, at this point, it seems likely that the full duplex efforts continue under the 802.11be TG umbrella. 

\item [\emph{d)}] \emph{Data and control plane separation.} 
802.11be devices with multi-band/multi-channel full duplex capabilities also have the unprecedented opportunity of separating the data and management planes \cite{2019PoKaiMultiLinkOp}. 
For instance, the STA buffer status feedback is currently performed using the same channel dedicated to data transmission and reception~\cite{2013NextGenWirelessLAN}, 
therefore introducing delays and overheads that translate into suboptimal scheduling decisions and throughput losses. 
These issues could be mitigated by dedicating a band/channel to data transmission/reception and a complementary one to provide frequent and reliable control information updates.
\end{itemize}

\vspace{-0.2cm}
\subsection{16\,spatial streams and multiple-input multiple-output (MIMO) protocol enhancements \label{sec:16SS}}

More antennas and better spatial multiplexing capabilities have been consistently added to Wi-Fi APs over the years to satisfy the stringent traffic demands generated by the increasing number of devices with wireless connectivity. For instance, 802.11ac APs can spatially multiplex up to eight spatial streams and four devices---in a MU-MIMO fashion---in a given downlink time/frequency resource, 
while 802.11ax enables APs to spatially multiplex up to eight single-stream devices in both downlink and uplink~\cite{2013NextGenWirelessLAN}. 

Consistent with this trend, 
and due to the 802.11be stringent requirements,
many Wi-Fi stakeholders foresee the need of further upgrading the APs' spatial multiplexing capabilities to accommodate for up to sixteen spatial streams~\cite{2018ChenDiscussionPHYFeatures, 2018HartEHfeatures}. This upgrade has the potential of doubling 802.11be spectral efficiency w.r.t. 802.11ax, 
taking full advantage of both 
the high speed backhaul provided by fiber-to-the-home (FTTH) solutions and 
the rich scattering in the indoor environments, 
where Wi-Fi systems typically operate.

Such spatial multiplexing gains, however, could be hindered by the overhead of the channel sounding process,
which is crucial to acquire accurate channel state information (CSI).
Doubling the number of spatial streams
while reusing the same explicit CSI acquisition procedure currently specified in 802.11ax may not be scalable. 
For this reason, 
802.11be is considering to introduce an implicit channel sounding procedure
that relies on STA-transmitted pilots 
and exploits uplink/downlink channel reciprocity~\cite{2018ChenDiscussionPHYFeatures}. 
Implicit sounding would likely require APs to implement a calibration method to prevent potential hardware mismatches that could break channel reciprocity.

\vspace{-0.2cm}
\subsection{Multi-access point coordination \label{sec:multi-AP}}

Enabling some degree of collaboration among neighbouring 802.11be APs will permit a more efficient utilisation of the limited time, frequency and spatial resources available,
and thus is also an appealing approach to enhance system performance.
Below we discuss three of the main alternatives considered within 802.11be, 
following an increasing order of coordination complexity.
\begin{itemize}

\item [a)] \emph{Coordinated OFDMA.} 
In coordinated OFDMA, 
collaborative APs synchronise their data transmissions, 
and use orthogonal time/frequency resources. 
This coordinated resource assignment diminishes the collision probability with respect to the case when APs implement independent contention-based channel access procedures \cite{2018ChenDiscussionPHYFeatures}.
Coordinated OFDMA is particularly attractive  to minimise the latency of short packet data transmissions, 
since it allows an efficient sharing and full occupation of the band by collaborating neighbouring devices, 
which otherwise would require multiple contention processes, 
and would not utilise the available resources up to their full potential. 

\begin{figure}[!t]
\centering
\includegraphics[width=5cm]{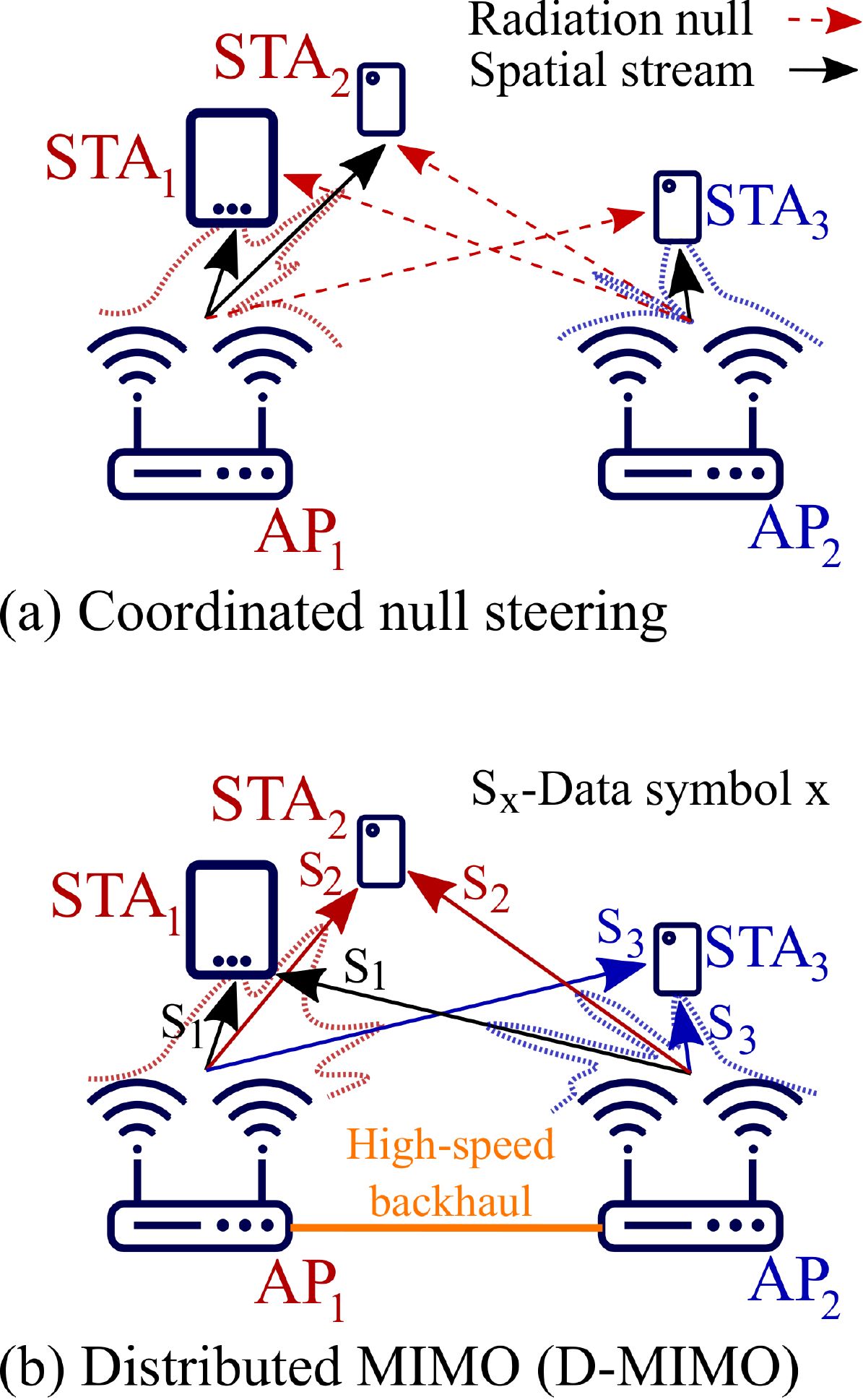}
\caption{(a) Coordinated null steering, and 
(b) distributed MIMO (D-MIMO) multi-AP coordination techniques.}
\label{fig:multiAP}
\vspace{-0.5cm}
\end{figure}


\item [b)] \emph{Coordinated Null Steering.} 
Multi-antenna APs typically use their capabilities for spatially multiplexing STAs in the same time/frequency resources, 
and/or to provide useful signal power gains through beamforming. 
Alternatively, APs can also leverage their antennas to place spatial radiation nulls from and towards non-associated devices in their neighbourhood, 
as shown in Fig.\ \ref{fig:multiAP}(a). 
This approach is referred to as coordinated null steering or coordinated beamforming, 
and is targeted at further boosting spatial reuse by enabling the simultaneous data transmission---using the same time/frequency resources---of devices within the same coverage area~\cite{2018ChenDiscussionPHYFeatures}. 
When compared to coordinated OFDMA, 
coordinated null steering generally requires a further degree of cooperation among overlapping basic service sets to organise scheduling decisions and facilitate the acquisition of CSI from non-associated devices, 
which is essential for the effective placement of radiation nulls.


\item [c)] \emph{Distributed MIMO (D-MIMO).} 
D-MIMO is the most intricate solution in terms of coordination complexity being considered by 802.11be. 
As illustrated in Fig.\ \ref{fig:multiAP}(b), D-MIMO non-collocated APs perform a joint data transmission and/or reception from multiple STAs reusing the same time/frequency resources~\cite{2018LopezDMIMOArchitectures, 2018PoratConstrainedDMIMO}. 
When compared with systems comprised of independent APs, 
the tight inter-AP collaboration of D-MIMO can provide an extended coverage, 
thanks to the additional beamforming gains, 
and an improved spatial multiplexing, 
as neighbouring APs are turned from interferers to servers~\cite{2018ChenDiscussionPHYFeatures}. 
A D-MIMO cluster is also more likely to use the up to 16 spatial streams that may be available in 802.11be. 
Achieving these gains requires inter-AP collaboration to jointly process both the data and the CSI of all STAs involved in the data transmission/reception, 
thus raising the need for a high-capacity, low-latency wired (e.g. fibre) or wireless (e.g. millimetre wave) backhauling network. 

Importantly, the implementation of D-MIMO in 802.11be would require the design of new distributed carrier-sense multiple access with collision avoidance (CSMA/CA) mechanisms, 
compliant with regulations, 
to optimise channel access and guarantee a fair coexistence with independent APs deployed in the same coverage area~\cite{2018LopezDMIMOArchitectures}. 
This aspect, together with the stringent time, frequency and phase synchronisation-related constraints imposed by Wi-Fi PHY layer numerology, 
have led a number of 802.11be contributions to propose D-MIMO implementations, 
consisting of a master AP that must be visible to all collaborating APs and oversees the cluster operation~\cite{2018PoratConstrainedDMIMO}. 
While theoretically compromising performance gains, 
the presence of a master AP could greatly simplify the coordination requirements, 
e.g. by enabling an efficient over-the-air synchronisation through the transmission of a control trigger frame to collaborating devices.

\end{itemize}


\vspace{-0.2cm}
\subsection{Enhanced link adaptation and retransmission protocol  \label{sec:linkAdaptation}}

Current Wi-Fi systems rely on the retransmission of MAC protocol data units (MPDUs) when these are not successfully decoded at the receiver or an acknowledgement (ACK) is not received at the transmitter. 
In this automatic repeat request (ARQ) approach, 
the receiver discards the failed MPDU before receiving its retransmitted version,
thus not allowing soft-combining. 
With the requirements on enhanced reliability and reduced latency, 
multiple companies advocate that 802.11be should evolve towards hybrid ARQ (HARQ) capabilities. 
HARQ-capable devices attempting to decode a retransmitted MPDU do not ignore the previous unsuccessful MPDU/s, 
but instead combine their soft-bits to improve the likelihood of correct decoding. 
The HARQ mechanism---already implemented in cellular systems and previously discussed during the standardisation of Wi-Fi\,5~\&~6---can provide signal-to-interference-plus-noise ratio (SINR) gains of approximately 4 dB with respect to ARQ in the ideal additive white Gaussian noise (AWGN) channel~\cite{2018HARQGainStudies}.

While the gains of implementing HARQ are clear in an ideal additive white Gaussian noise channel, 
some Wi-Fi stakeholders have reservations regarding the potential throughput gains achieved by HARQ when considering that Wi-Fi scenarios frequently suffer from bursty interference due to collisons~\cite{2018HartEHfeatures}. 
Additionally, from a computational standpoint, 
the soft combining operation of HARQ requires extra computational capabilities as well as memory requirements for storing past transmissions. 
At this stage, further studies are expected to evaluate the performance and complexity requirements of HARQ in Wi-Fi scenarios/use cases. 


\vspace{-0.2cm}
\section{Coexistence in the 6\,GHz band \label{sec:coexistence}}

802.11be is eager to make full use of the 1.2\,GHz of spectrum potentially available in the 6\,GHz band, 
as discussed in Sections~\ref{sec:320MHz} and~\ref{sec:multi-channel}.\footnote{We emphasize that, at the time of writing this article, it is plausible that regulators decide not to dedicate the entire 6\,GHz band to unlicensed usage, and that this decision may vary per country.}
However, to access these enticing frequency resources, 
Wi-Fi will have to coexist with other technologies operating in the same band. 
Two major types of technologies are foreseen to share the 6\,GHz band:
\begin{itemize}
\item
Existing---and to be deployed---fixed and mobile services, 
which can be considered as incumbents, and
\item
Newcomers, 
where we can distinguish between IEEE-based technologies, 
such as 802.11ax or 802.11be, 
and 3GPP-based ones, 
such as New Radio-Unlicensed (NR-U).
\end{itemize}

Based on this classification, 
it is likely that regulatory bodies define new coexistence requirements to guarantee that newcomers do not generate harmful interference to incumbent services in the 6\,GHz band, 
as shown in Fig.\ \ref{fig:FCCAllocation}~\cite{FCCUnlicensed6GHz}. 
For instance, in the USA, it seems sensible that incoming technologies guarantee a peaceful coexistence with the more than 44.000 fixed access links deployed today, 
since the latter represent an important component of current and future cellular technologies, 
where they mostly serve as a backhaul solution. 

While some coexistence schemes have been already proposed by several stakeholders, 
such as that in~\cite{2018RFKengineering2}---based on a proactive geolocation and database-based approach as well as some a posterior interference detection and mitigation techniques---the definition of the coexistence technique of choice is still under discussion within the relevant regulatory bodies at the time of writing this article~\cite{FCCUnlicensed6GHz}. 
In this regard, it is worth noting that some regulators have already specified coexistence techniques that may also be of value for the 6\,GHz operation, 
e.g. the dynamic frequency selection method applied in the 5\,GHz band, the geo-location database-based approach used in the television white spaces, or the spectrum access system (SAS) employed in the citizens broadband radio service (CBRS)~\cite{FCCUnlicensed6GHz}.

In contrast, coexistence between new comers, 
such as 802.11ax/be and NR-U, 
is likely to be governed by listen-before-talk~\cite{2018IeeeStandardAxDraft}. 
However, and differently to the 5\,GHz coexistence case, 
802.11ax/be may not be treated as an incumbent technology, 
and thus advantages in terms of energy detection threshold may not be given. 
Within this space, 
the use of a common preamble among multiple technologies to realise a cross-technology preamble detection and/or virtual carrier sense-like mechanism has also been considered to enhance coexistence. 
Similarly, interested stakeholders are evaluating potential issues that latency sensitive traffic such as Wi-Fi beacons and NR-U discovery reference signals (DRSs) may experience when coexisting technologies implement listen-before-talk processes with different durations.
These and other inter-technology aspects are already under discussion, 
and is expected that the IEEE and the 3GPP can leverage their previous experiences in the 5\,GHz band to smoothly find efficient coexistence solutions.

\vspace{-0.2cm}
\section{802.11be Performance Evaluation}
\label{sec:simResults}

In this section, 
we present the results of detailed system-level simulations
performed on a typical enterprise scenario,
to assess the actual throughput gains that 802.11be may bring over 802.11ax
under realistic conditions.
Specifically, 
we adopt the view of a company residing in a $40$\,m $\times 40$\,m building of 3\,m height,
which aims at upgrading their $16$ 802.11ax APs---deployed in a square grid fashion and reusing 4 channels---with newer APs that implement a key subset of the 802.11be technical features introduced in Sec.~\ref{sec:features}, 
namely:
\begin{enumerate}
\item  \emph{More bandwidth:} 
802.11be-capable APs support 160\,MHz transmissions in the 6\,GHz band, 
a feature likely to be mandatory in 802.11be, 
whereas 802.11ax APs perform 80\,MHz transmissions.

\item  \emph{More antennas and spatial streams per AP:} 
802.11be-capable APs incorporate $16$ antennas,
and can spatially multiplex up to $16$ STAs in downlink and uplink---doubling the number of antennas and spatial streams w.r.t. 802.11ax APs.

\item \emph{Implicit CSI acquisition:} 
802.11be-capable APs rely on STA-transmitted pilots to estimate the channel. 
This allows 802.11be to reduce the overhead introduced by the explicit CSI acquisition procedure of 802.11ax.\footnote{
For simplicity and to facilitate the comparison, 
we consider that there are no CSI acquisition errors in both 802.11be and 802.11ax systems.}
\end{enumerate}
A detailed list of the most relevant system parameters can be found in Table~\ref{table:parameters}. The results have been averaged over 100 random simulation drops,
where each drop simulates 10\,s of operation.

\begin{table}
\centering
\caption{Detailed system-level parameters}
\vspace{-0.15cm}
\label{table:parameters}
\def\arraystretch{1.1}
\begin{tabulary}{\columnwidth}{ |p{3.85cm} | p{4.05cm} | }
\hline
\rowcolor{LightBlue}
\textbf{Parameter} 		& \textbf{Description} 						\\ \hline
\rowcolor{LightCyan}
\textbf{Deployment} 		&  							\\ \hline
Floor size			& $40~\textrm{m}\times 40~\textrm{m}$					\\ \hline
AP positions			& 16 ceiling-mounted APs  equally spaced ($d_{x} = d_{y} = 10$\,m)		\\ \hline
AP/STA heights		& $h = 3/1$\,meters						\\ \hline
STA distribution 		& 512 uniformly deployed STAs. $10$\,cm of min.\ inter-STA distance. 					\\ \hline
AP-STA association criterion & Strongest average received signal 	
\\ \hline
\rowcolor{LightCyan}
\textbf{PHY \& MAC} 		&  								\\ \hline
Carrier frequency		& 5.18\,GHz (.11ax) / 6.2\,GHz (.11be)						\\ \hline
Total system bandwidth 		& 320\,MHz (.11ax) / 640\,MHz (.11be)						\\ \hline
Channel size 		& 80\,MHz (.11ax) / 160\,MHz (.11be)						\\ \hline
OFDM guard interval duration 	& $0.8\mu s$ 							\\ \hline
AP/STA maximum TX power	& $P_{\textrm{max}} = 24/15$\,dBm						\\ \hline
Number of antennas per AP 	& 8 in a $4\times 2$ planar array (.11ax) / 16 in a $4\times 4$ planar array (.11be)	\\ \hline
Number of antennas per STA 	& 1								\\ \hline
AP and STA antenna elements      	& Omnidirectional with 0 dBi and $0.5\lambda$ separation 				\\ \hline
CCA energy detection threshold 	& $\gamma_{\textrm{LBT}} = -62$ dBm 				\\ \hline
Signal detection threshold		& $\gamma_{\textrm{preamble}}=-82$~dBm with -0.8 dB of minimum SINR 		\\ \hline
MCS selection algorithm 		& Minstrel 							\\ \hline
AP/STA noise figure 		& $F_{\textrm{dB}} = 7/9$\,dB					\\ \hline	 
Maximum \# of scheduled STAs 	& 8 (.11ax) / 16 (.11be)						\\ \hline
AP spatial precoding/detection & Zero Forcing							\\ \hline
STA scheduling 		& Round Robin with semi-orthogonal user selection (SUS)			\\ \hline
Downlink power allocation 	& Equal power assigned per STA					\\ \hline
MPDU payload size		& 1500\,bytes 						\\ \hline
Medium access policy & Distributed coordination function with AP-scheduled uplink access 				\\ \hline
Maximum TXOP length		& 4\,ms 							\\ \hline
\rowcolor{LightCyan}
\textbf{Channel model} 		&  							\\ \hline
Path loss and LOS probability	& 3GPP 3D InH (3GPP TR 38.901) 					\\ \hline
Shadowing 			& Log-normal with $\sigma = 3/8 $ dB (LOS/NLOS) (3GPP TR 38.901) 		\\ \hline
Fast fading 			& Ricean with log-normal K factor (3GPP TR 38.901)  			\\ \hline
Thermal noise 		& -174 dBm/Hz spectral density					\\ \hline
\rowcolor{LightCyan}
\textbf{Traffic model} 		&  							\\ \hline
Traffic model			& FTP model 3 with a packet size of 0.5\,MBytes				\\ \hline
Traffic generated per STA		& 75\,Mbits/s							\\ \hline
Downlink/Uplink traffic ratio		& 0.5/0.5							\\ \hline	 
\end{tabulary}
\vspace{-0.7cm}
\end{table}

\begin{figure}[!t]
\centering
\includegraphics[width=8.25cm]{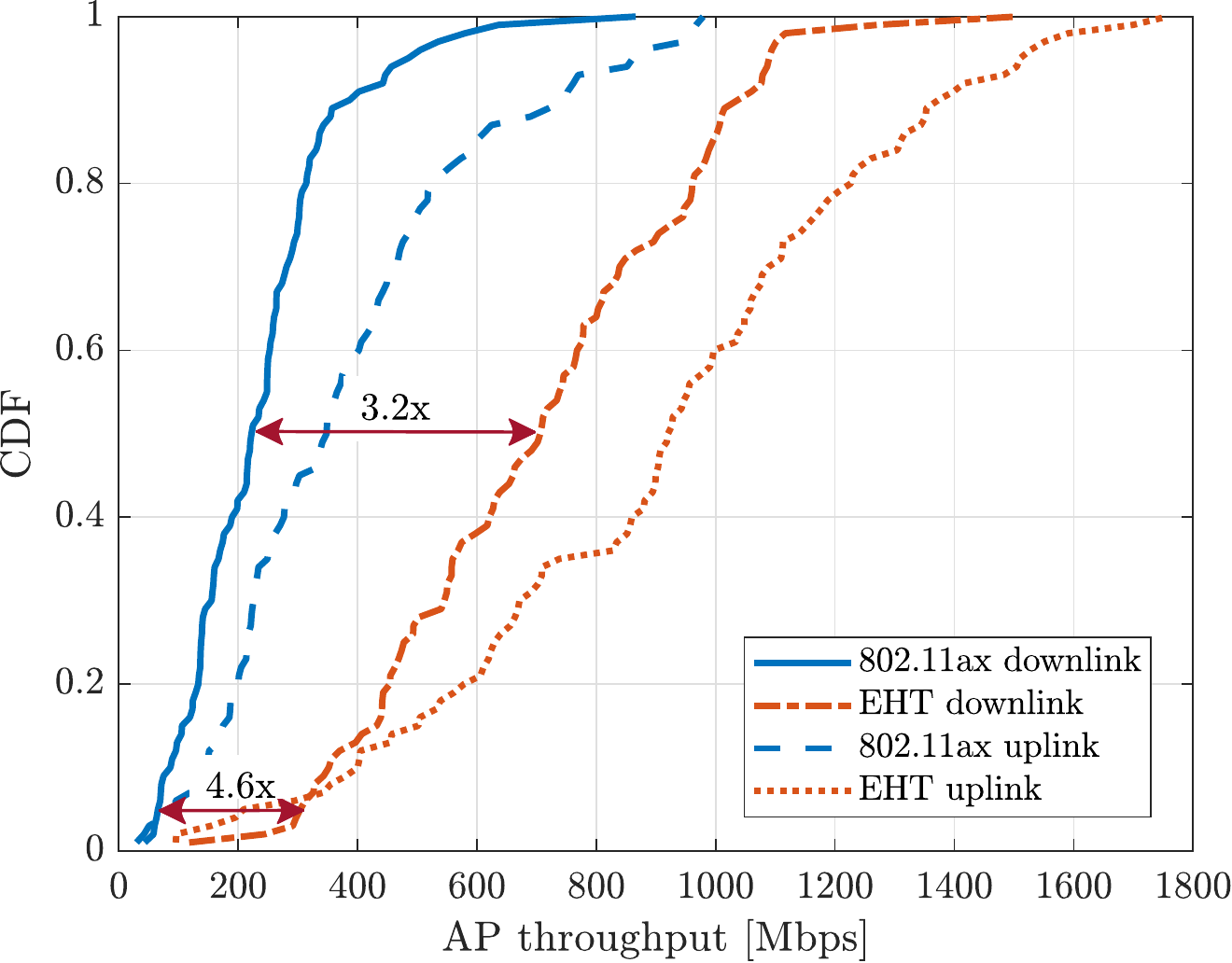}
\caption{CDF of the downlink and uplink aggregate throughputs per AP [Mbps] in the considered 802.11ax and 802.11be enterprise scenarios.}
\label{fig:throughputResults}
\vspace{-0.5cm}
\end{figure}

Fig.\ \ref{fig:throughputResults} shows the cumulative distribution function (CDF) of the aggregate downlink and uplink throughput per AP measured at the MAC data service access point for both 802.11ax and 802.11be deployments. 
In this enterprise scenario, 
it can be observed that STAs generally experience smaller throughputs in the downlink than in the uplink despite of the identical downlink/uplink traffic generation ratio,
which is a consequence of 
\begin{itemize}
\item [\emph{a)}] the downlink power division that APs perform in downlink when realising spatial multiplexing,
\item [\emph{b)}] the non-existing beamforming capabilities at the single-antenna STAs---which cannot reject interference through a receive spatial filter---, and
\item [\emph{c)}] the larger noise figure at the less complex and cheaper STAs---which leads to reduced downlink SINRs.
\end{itemize}
It can also be observed that the gap between the downlink and uplink performance slightly grows with the number of antennas at the AP, i.e. between 802.11ax and 802.11be deployments. 
This is because the impacts of the downlink power division---mentioned in \emph{a)}---and the uplink interference rejection---mentioned in \emph{b)}---become more significant when APs are equipped with a larger number of antennas, therefore relatively enhancing more the uplink throughput than the downlink one.


As envisioned, 
Fig.\ \ref{fig:throughputResults} demonstrates that 802.11be provides notable throughput gains w.r.t.\ 802.11ax, 
in more detail, $3.2\times$ and $2.7\times$ in median for downlink and uplink, respectively, and $4.6\times$ and $2.2\times$ in 5\%-tile downlink and uplink throughputs, respectively.
This is thanks to the larger transmission bandwidths and enhanced spatial multiplexing capabilities of 802.11be. 
Interestingly, the throughput gains provided by 802.11be in this realistic scenario do not always reach the maximum theoretical gain of $\approx 4\times$ w.r.t.\ 802.11ax because 
\emph{a)} cell-edge STAs might not actually benefit from the larger bandwidth available in 802.11be, 
since their SINRs---and, consequently, their modulation and coding schemes (MCSs)---diminish due to the larger noise power, 
which grows proportionally with the transmission bandwidth, 
\emph{b)} in spite of generally scheduling a larger number of STAs than 802.11ax APs, 
802.11be APs are not always able to spatially multiplex $16$ STAs, and 
\emph{c)} the per-STA SINRs generally degrade when more STAs are spatially multiplexed, 
since the spatial correlation of their channels increases.

To reach the maximum theoretical gain of $\approx 4\times$ that 802.11be can bring w.r.t.\ 802.11ax---thanks to doubling the bandwidth and the spatial streams---some degree of coordination between APs will be required, 
urging for the need of examining further such techniques in 802.11be. Further studies are also needed to understand 802.11be gains in the presence of coexisting 802.11 legacy devices performing, e.g., contention-based access. 

\vspace{-0.2cm}
\section{Conclusion\label{sec:Conclusion}}

In this paper, 
we have presented a comprehensive overview of the initial steps taken towards the creation and standardisation of 802.11be---the next generation Wi-Fi beyond 802.11ax.
In more detail, we have covered 802.11be main objectives and expected timelines, 
shared the viewpoints of different Wi-Fi stakeholders,  and discussed its main candidate features---providing insights on their benefits and challenges as well as system-level simulation results in a typical enterprise scenario. 
The 802.11be standardisation process has just started
and everything is still open,
please come and join us in making a better Wi-Fi.


\vspace{-0.5cm}
\bibliographystyle{IEEEtran}
\bibliography{Ming_library}

\end{document}